\documentclass[10pt,leqno]{amsart}
\usepackage{graphicx}
\usepackage{indentfirst,csquotes}
\usepackage{amssymb,amsthm,amsmath}
\usepackage{xcolor,paralist,fancyhdr,etoolbox}
\usepackage{hyperref}
\usepackage{nomencl}
\usepackage{mdframed}
\usepackage{mathptmx}

\hypersetup{ colorlinks=true, linkcolor=black, filecolor=black, urlcolor=black }

\usepackage{lipsum}
\newenvironment{acknowledgments}{%
  \section*{Acknowledgments}%
}{}

\begin{document}

\title{The impact of pressure oscillations on bubble rising in shear-thinning fluids} 
\author[Mario Riccio]{Mario Riccio}
\author[Marco De Corato]{Marco De Corato}
\date{\today}
\email{mdecorato@unizar.es}
\maketitle

\let\thefootnote\relax
\footnotetext{MSC2020: Primary 00A05, Secondary 00A66.} 

\begin{abstract}
We study the rising dynamics of a bubble driven into periodic volumetric oscillations by an external pressure driving within a highly viscous shear-thinning fluid. We perform axisymmetric direct numerical simulations employing the Carreau-Yasuda model to describe the rheological behavior of the fluid and the finite element method to discretize the equations. We carry out a parametric study of the bubble rising dynamics, changing the amplitude and the frequency of the external pressure driving, and the bubble radius. Due to the external pressure oscillations, the bubble undergoes volume changes that strain the liquid at much larger rates than those due to natural rising, causing the surrounding fluid viscosity to thin. 
The numerical results show that the rising dynamics become highly nonlinear and unsteady due to the interplay of the shear-thinning rheology and the external driving. As a result, the period-averaged rising velocity of the bubble can increase by orders of magnitude compared to its natural rising velocity. These nonlinear effects become progressively more important as the amplitude and the frequency of the pressure driving or the bubble radius are increased. 
Qualitatively, the simulation model agrees with previous experimental findings in terms of average rising velocity. However, the experiments exhibit terminal velocities that are smaller than those predicted numerically, along with differences in bubble shape during the ascent. These discrepancies may be attributed to modeling the fluid rheology as a generalized Newtonian fluid rather than as a viscoelastic one.

\end{abstract}

\maketitle

\section{\label{sec:level1}Introduction}

The controlled release and transport of gas bubbles in shear-thinning liquids underpin processes such as polymer foaming\cite{Taki2006BubCoal}, microfluidic reactors\cite{garstecki2005formation}, food manufacturing like ice cream\cite{sofjan2004effects}, and personal-care formulations\cite{lin1970mechanisms}. Gas bubbles are virtually omnipresent in every industrial application where fluid flow is required \cite{rodriguez2015generation,lohse2018bubble,garbin2025bubbles}. In many of these industrial processes, bubbles can serve as an integral part of the product, to improve, for instance, the texture \cite{luyten2004crispy} or emerge undesirably, bringing bacterial contamination \cite{lin1970mechanisms}, making the study of degassing mechanisms essential for optimizing growth and removal.

Beyond engineered systems, bubbles play a role in geophysical contexts, particularly in volcanic eruptions, \cite{seropian2021review}. Volatile species dissolved in silicate melts degass as pressure decreases during magma ascent, forming bubbles that dictate eruption style and intensity \cite{gardner2023bubble}. Such bubbles drive buoyancy, fragmentation, and explosive transitions.
Seismic waves or pressure oscillations can trigger eruptions by modulating bubble behavior. Sloshing in bubbly magma reservoirs induced by earthquakes that bring to pressure oscillations causes foam collapse in layered systems, releasing gas, enhancing heat transfer, and promoting magma mixing for delayed eruptions \cite{namiki2016sloshing}.

The problem of how bubbles affect fluids is also investigated in the construction field, where bubble removal from structural materials like cement is key, as their presence can degrade mechanical properties. In fresh cement mortar, increasing the proportion of air bubbles in the 100–500 $\mu m$ range significantly reduces fluidity, while yield stress exhibits a correlation with fluidity at constant air content \cite{guo2022characteristic}.

Unlike Newtonian fluids, where classical mathematical correlations exist for terminal bubble rise velocities \cite{clift1978bubbles}, complex fluids lack predictive models. Indeed, complex fluids can exhibit non-constant viscosity that decreases with increasing shear rate, display viscoelastic normal stresses, or even a yield stress. In such media, bubbles on the millimeter-to-centimeter scale undergo dynamic phenomena that are still not fully explored \cite{dollet2019bubble} \cite{de2019rising}\cite{zhang2023drag}.

Given its relevance in several industrial applications, it is no surprise that the rising of bubbles in complex fluids has a long history of intense research \cite{astarita1965motion,hassager1979negative}. 
Over the past two decades, a combination of high-speed experiments and detailed simulations has revealed how the non-Newtonian fluid rheology impacts the rising of bubbles. Pillapakkam \textit{et. al} demonstrated that rising bubbles in viscoelastic polymer solutions develop cusp-shaped rear profiles and exhibit a six-fold velocity jump at a critical size. The shape of the bubble and of its wake structure changes fundamentally at the onset of the jump \cite{Pillapakkam2007}.
They also proposed a universal correlation of non-dimensional numbers for the dimensionless critical bubble volume at the jump discontinuity \cite{PILZ2007124}. Subsequently, a computational and theoretical study by Fraggedakis \textit{et. al} investigated the critical bubble volume in viscoelastic fluids. They found that the velocity discontinuity is due to a hysteresis loop, arising from the formation of a tip at the rear pole. They also found the characteristic negative wake effect, already observed by Hassager\cite{hassager1979negative}.
Their predictions achieved quantitative agreement with the experimental results of Pilz and Brenn, as well as qualitative agreement with larger bubbles involving inertia \cite{fraggedakis2016velocity}. 

The rising of millimiter to micrometer-sized bubbles in complex fluids can be extremely slow due to their large viscosity or even impossible if the bubble buoyancy does not overcome the yield stress \cite{tsamopoulos2008steady,moschopoulos2021concept,pourzahedi2022flow,esposito2024buoyancy}.
One promising approach to remove these bubbles involves enhancing their rising velocity through periodic pressure changes or vibrations.
The phenomenon of degassing by acoustic cavitation has been studied for several years \cite{eskin2017overview}. This approach harnesses violent bubble cavitation under strong acoustic fields. Instead, in the case of complex fluids, stimuli at frequencies and amplitudes much lower than that of the cavitation threshold can locally alter viscosity around the bubble or cause the material to yield. This strategy directly leverages the mechanical stress generated by oscillating bubbles to facilitate bubble release. Additionally, ultrasound-induced secondary Bjerknes forces have been suggested as a way to encourage bubble merging, thereby increasing their size and ascent speed \cite{pelekasis2004secondary}.
In contrast to chemical antifoaming agents, the application of these mild mechanical stimuli leaves no residual additives, preserving purity and rheology while eliminating concerns about fouling or compatibility.

Stein \textit{et. al} first showed that low-frequency pulsations enhance bubble release from yield-stress fluids. They showed that small bubbles trapped in carbopol gels can be released by applying pressure oscillations \cite{stein2000rise}. Subsequent works shed more light on the interplay between pressure oscillations, bubble dynamics and the fluid yield stress \cite{karapetsas2019dynamics,de2019oscillations,saint2020acoustic}. Nevertheless, a complete understanding of how the high-speed bubble dynamics introduces plastic deformations and finally escapes the yield-stress fluid is missing.

More closely related to the topic of this work, Iwata \textit{et al.} showed that the application of pressure oscillations also strongly enhances the rising speed of millimetric bubbles in viscoelastic shear-thinning fluids \cite{iwata2008pressure}. In their work, they found that mild pressure oscillations could increase the rising speed by up to three orders of magnitude compared to their natural rising speed. They hipothesized that the fast radial excursions of the bubble introduce large rate of strains in the fluid leading to significant shear thinning. This hypothesis is supported by their subsequent study where they use PIV to observe that the bubble develops a negative wake \cite{hassager1979negative}, which is a signature of shear thinning effects \cite{iwata2019local}.

 To assess the importance of shear thinning, we investigated the drag force experienced by a slowly pulsating bubble \cite{de2019rising}. In this model, we assumed low-Reynolds numbers, we fixed the kinematics of the bubble and we employed the Carreau–Yasuda model to describe the fluid rheology. We found that the radial oscillations significantly decreased the bubble drag. However, the effect was larger than that observed in the experiments. 
 
 Zhang \textit{et. al} solved the unsteady Stokes problem for a spherical bubble translating through a shear-thinning fluid at a constant velocity plus a high-frequency sinusoidal oscillation. They find a transition from quadratic to power-law scaling between drag reduction and oscillation amplitude \cite{zhang2023drag}.

Despite their valuable insights, these theories retain restrictive assumptions: fixed spherical shape, negligible inertia, or a clear separation of pulsation and translation timescales.
Such assumptions are removed in the present work, which addresses a key research gap left by previous studies. 
Here, we study bubble rising in a shear-thinning fluid under an externally-applied oscillatory pressure by solving the fully coupled Navier–Stokes equations for a Carreau–Yasuda fluid. We track the deformable gas–liquid interface of the bubble using an arbitrary Lagrangian–Eulerian (ALE)\cite{donea2004arbitrary} formulation and account for fluid inertia. This enables the model to capture transient inertial effects and arbitrary bubble deformation.

\makenomenclature
\begin{mdframed}[linewidth=0.5pt, roundcorner=0pt, skipabove=10pt, skipbelow=10pt]
\printnomenclature[1cm]

\nomenclature[a01]{$R_c$}{Radius of the computational domain [m]}
\nomenclature[a02]{$R_0$}{Initial bubble radius [m]}
\nomenclature[a03]{$\Gamma_1$}{Bubble Surface}
\nomenclature[a04]{$\Gamma_2$}{Outer domain surface}
\nomenclature[a05]{$\Omega$}{Fluid Domain}
\nomenclature[a06]{$\mathbf{n}$}{Normal surface vector}
\nomenclature[a07]{$V_0$(t)}{Bubble volume at the initial time [m$^3$]}
\nomenclature[a08]{$V_b$(t)}{Bubble volume in time [m$^3$]}
\nomenclature[a09]{$\kappa$}{twice the local mean curvature [1/m]}

\nomenclature[b01]{$\rho$}{Density [kg/m$^3$]}
\nomenclature[b02]{$\sigma$}{Surface Tension [N/m]}
\nomenclature[b03]{$\eta_0$}{Zero-shear Viscosity [$Pa\cdot s$]}
\nomenclature[b04]{$\eta_\infty$}{Viscosity at an infinite shear rate [$Pa\cdot s$]}
\nomenclature[b05]{$\eta$}{Shear-dependent viscosity [$Pa\cdot s$]}
\nomenclature[b06]{$\lambda$}{Relaxation time parameter in the Carreau–Yasuda model [s]}
\nomenclature[b07]{$n$}{Power index in the Carreau–Yasuda model}
\nomenclature[b08]{$a$}{Model parameter in the Carreau–Yasuda model}

\nomenclature[c01]{$\hat{r}$}{Radial unit vector}
\nomenclature[c02]{$\hat{z}$}{Axial unit vector}
\nomenclature[c03]{$\mathbf{r}$}{Position vector}
\nomenclature[c04]{$t$}{Time [s]}
\nomenclature[c05]{$\mathbf{v}$}{Velocity [m/s]}
\nomenclature[c06]{$z_{cm}$}{Center of mass position [m]}
\nomenclature[c07]{$\mathbf{D}$}{Rate-of-strain tensor [s$^{-1}$]}
\nomenclature[c08]{$\dot{\gamma}$}{Second invariant of $\mathbf{D}$ [s$^{-1}$]}
\nomenclature[c09]{$v_0$}{Natural rising velocity [m/s]}
\nomenclature[c10]{$v_\text{Stokes}$}{Natural rising velocity evaluated with Stokes Law [m/s]}

\nomenclature[d01]{$p$}{Pressure [Pa]}
\nomenclature[d02]{$p_B$}{Bubble internal pressure [Pa]}
\nomenclature[d03]{$p_A$}{Atmospheric pressure [Pa]}
\nomenclature[d04]{$g$}{Gravitational acceleration [m/s$^2$]}

\nomenclature[e01]{$f$}{Frequency of the oscillating pressure [Hz]}
\nomenclature[e02]{$T$}{Period of the oscillating pressure [s]}
\nomenclature[e03]{$k$}{Dimensionless amplitude of the oscillating pressure [-]}

\nomenclature[f01]{G$_b$}{Dimensionless acceleration [-]}

\nomenclature[g01]{$R$e}{Reynolds number}
\nomenclature[g02]{$Ca$}{Capillary number}
\nomenclature[g03]{$Cu$}{Carreau number}
\nomenclature[g04]{$Cu_\omega$}{Carreau number based on frequency}
\nomenclature[g05]{$Bo$}{Bond number}
\nomenclature[g06]{$Wo$}{Womersley number}
\nomenclature[g07]{$\langle \rangle$}{quantity averaged over one period}

\end{mdframed}

\section{Governing equations}
\begin{figure}[h]
\includegraphics[width=0.275\textwidth]{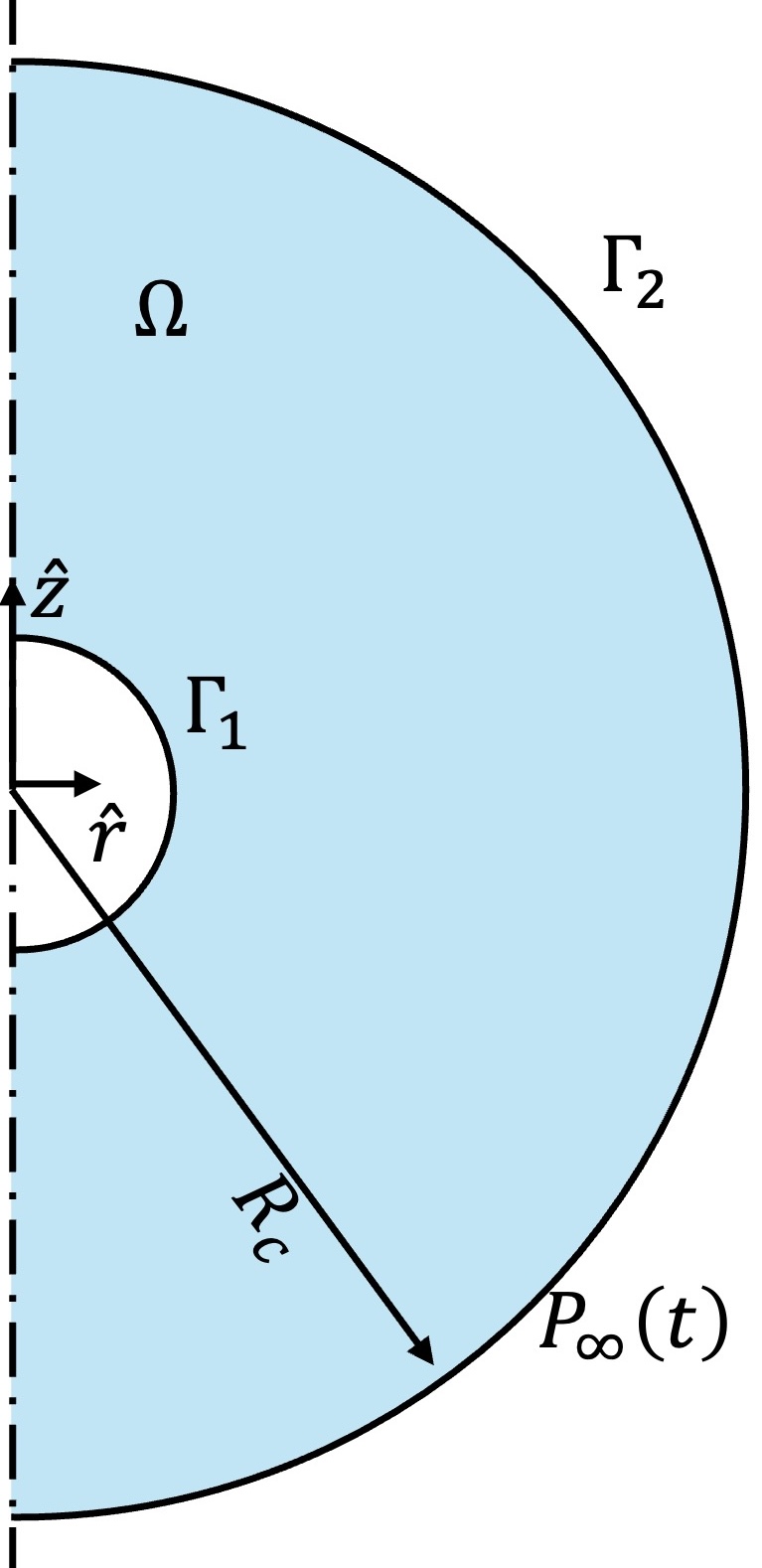}
\caption{Sketch of the axisymmetric computational domain used in this study. A fixed cylindrical reference frame is employed. The bubble is initially a sphere of radius $R$ placed at the origin and is suspended in a shear-thinning fluid. A periodic disturbance to the hydrostatic pressure is applied at the outer boundary $\Gamma_2$, which drives volumetric oscillation of the bubble. }
\label{Problem_Scketch}
\end{figure}

Motivated by the experiments of Iwata \textit{et al.} \cite{iwata2008pressure}, we study in the rising dynamics of a bubble suspended in a shear-thinning fluid where the far-field pressure oscillates periodically in time. We employ transient simulations to study the impact of periodic pressure oscillations on the bubble shape and position as it rises through the shear-thinning fluid. As shown schematically in Figure \ref{Problem_Scketch}, we consider an axisymmetric bubble suspended in an incompressible fluid domain, $\Omega$, subjected to a gravitational acceleration $g$. We assume that the problem is isothermal and that no mass transfer occurs between the gas and the fluid. We employ a fixed cylindrical reference frame (see Figure \ref{Problem_Scketch}). Since the fluid is incompressible, the velocity field, $\mathbf{v}$, satisfies
\begin{equation}\label{cont_eq}
\nabla \cdot \mathbf{v} \;=\; 0.
\end{equation}
We assume that the fluid density, $\rho$ is constant and that the shear-thinning behavior of the viscosity can be captured by the  Carreau-Yasuda model. As a result, the momentum balance reads
\begin{equation}
\rho \,\left( \frac{\partial \mathbf{v}}{\partial t} 
\;+\; \mathbf{v}\cdot\nabla\mathbf{v} \right) 
\;=\; -\nabla p 
\;+\; \nabla \cdot \Bigl[2\,\eta(\dot{\gamma})\,\mathbf{D}\Bigr]- g \rho \hat{\boldsymbol{z}},
\end{equation}
where $\hat{\boldsymbol{z}}$ is the unit vector directed along the z-axis (see Figure \ref{Problem_Scketch}), $p$ is the pressure, and $\mathbf{D}$ is the rate-of-strain tensor defined as 
\begin{equation}
\mathbf{D} = \frac{1}{2}\left(\nabla \mathbf{v}+ \nabla \mathbf{v}^T  \right) \, \, .
\end{equation}
The effective viscosity $\eta(\dot{\gamma})$ follows the Carreau--Yasuda model:

\begin{equation}
\eta(\dot{\gamma}) 
\;=\; \eta_{\infty} \;+\; (\eta_{0} - \eta_{\infty}) 
\Bigl[\,1 + (\lambda \,\dot{\gamma})^{\,a}\Bigr]^{\tfrac{n-1}{a}},
\end{equation}
in which $\eta_{0}$ is the zero-shear viscosity, $\eta_{\infty}$ is the infinite-shear viscosity, $\lambda$ is a characteristic time constant, $a$ is the transition steepness parameter, and $n$ is the power-law exponent. The quantity $\dot{\gamma}$ is taken to be the second invariant of the rate-of-strain tensor, often expressed by
\begin{equation}
\dot{\gamma} 
\;=\; \sqrt{2\,\mathbf{D} : \mathbf{D}}.
\end{equation}

The governing equations must be supplemented with appropriate boundary and initial conditions. At the surface of the bubble, $\Gamma_1$ in Figure \ref{Problem_Scketch}, the balance of normal stresses reads: 
\begin{equation}\label{bc1}
[-\,p\,\mathbf{I} \;+\; 2\,\eta(\dot{\gamma})\,\mathbf{D}]\cdot\mathbf{n} 
\;=\; -\,p_{B}(t)\,\mathbf{n} \;+\; \sigma \,\kappa\,\mathbf{n},
\end{equation}
where $\mathbf{n}$ is the outward unit normal vector to the bubble surface, $\sigma$ is the (constant) surface tension coefficient, $\kappa$ is the local mean curvature of the gas-fluid interface. In the boundary condition, Eq. \eqref{bc1}, $p_{B}(t)$ is the pressure inside the bubble for which a compression law must be specified. Since we assumed an isothermal system, the pressure is given by an isothermal compression law
\begin{equation}
p_{B}(t) \;=\; p_B(0) \,\frac{V_{0}}{V_{B}(t)},
\end{equation}
where $V_{B}(t)$ is the bubble volume at time $t$, and $p_B(0)$ and $V_{0}$ are the gas pressure and bubble volume at time $t=0$, respectively. Using the divergence theorem, the volume of the bubble at any time can be expressed as a surface integral
\begin{equation}\label{vol_bubbl}
V_B(t) = \frac{1}{3} \int_{\Gamma_1} \mathbf{r} \cdot \mathbf{n} \, \, d\Gamma_1 \, \, ,
\end{equation}
with $\mathbf{r}$ denoting the distance from the origin of the reference frame to a point on the bubble surface. Likewise, the instantaneous z-coordinate of the bubble center of mass, $z_{\text{cm}}$, is computed as a surface integral
\begin{equation}
    z_{\text{cm}}=\frac{1}{4 V_{B}(t)} \int_{\Gamma_1} z \, \mathbf{r} \cdot \mathbf{n} \, \, d\Gamma_1 \, \, .
    \label{eq:pos}
\end{equation}
The initial bubble pressure is assumed to be given by the Young-Laplace equation $p_B(0) = 2\sigma / R + p_A $ with the atmospheric pressure $p_A = 101.325 \rm{kPa}$. 
On the axis os symmetry, the radial velocity, $v_r$, and its radial derivative are zero, $v_r=0$ and $\partial v_r/ \partial r=0$.

Far from the bubble, an open boundary condition is used on $\Gamma_2$. On this boundary, we fix the pressure as the sum of the initial gas pressure, the hydrostatic pressure,e and a sinusoidal disturbance 
\begin{equation}
p(t,z)=p_A  \left[1+ k  \sin{(2\, \pi f t)} \right] - \rho g z\, \, ,
\label{press_driv}
\end{equation}
with $k$ and $f$ the dimensionless amplitude and frequency of the disturbance. We then define the forcing period as $T=1/f$ This condition is implemented on the boundary of the computational domain, which has a radius $R_c = 200 \, R$ (See Figure \ref{Problem_Scketch}). We also need to fix the z-component of the velocity in our domain. Here we fix the $v_z=0$ at the point $r=R_c$ and $z=0$\cite{karapetsas2019dynamics}. The computational boundary, $\Gamma_2$ is placed sufficiently far away from the bubble that its impact on the dynamics of the bubble is negligible. 
Finally, at time $t=0$, the bubble is spherical with an initial radius $R$, an initial volume $V_{0}=4/3 \pi R^3$, an initial position $z(0)=0$, and the fluid surrounding the bubble is quiescent. 

To model a system close to the experiments\cite{iwata2008pressure}, we study bubbles with a millimetric initial radius $R \approx 1 \rm{mm} $ driven by pressure oscillation frequencies in the range $f=1-300 \rm{Hz}$. Since in the experiments the amplitude of the pressure driving couldn't be measured, we investigate a broad range of $k \in [0.05, 0.75]$, which represent amplitudes ranging from 5 to 75 kPa. The remaining physical and rheological parameters are summarized in Table I. Briefly, we assume that the density is that of water and the surface tension is that of a clean fluid-air interface. The rheological parameters are chosen to capture the shear-thinning behavior of the aqueous sodium polyacrylate (SPA) solution used in the experiments. The SPA solution that Iwata et. al.\cite{iwata2008pressure} displays a very large zero-shear viscosity and a strong shear-thinning behavior. 

\begin{table}[h!]
\caption{\label{tab:parameters} Fluid parameters used in the simulations}
\centering
\begin{tabular}{cc}
\hline
Physical property & Value and unit  \\
\hline
$\rho$ & $1003.5~\mathrm{kg/m^3}$ \\
$\sigma$ & $0.072~\mathrm{N/m}$ \\
$\eta_0$ & $90~\mathrm{Pa\cdot s}$ \\
$\eta_{\infty}$ & $10^{-3}~\mathrm{Pa\cdot s}$ \\
$a$ & 2 \\
$n$ & 0.3 \\
$\lambda$ & 57 s \\
\hline
\end{tabular}
\end{table}

To solve the governing equations \eqref{cont_eq}-\eqref{press_driv}, we use the finite element method that we developed and validated previously.
The finite element method has been widely employed to study the dynamics of bubbles rising in complex fluids \cite{fraggedakis2016velocity}.
To fulfill the inf-sup condition, we use quadratic interpolation functions for the velocity field and linear interpolation functions for the pressure.
We discretize the axisymmetric computational domain, $\Omega$ in Figure \ref{Problem_Scketch}, into triangular elements with a more refined mesh close to the bubble surface (See Figure \ref{fig:mesh_refinement}). To handle the mesh deformation due to the movements of the bubble surface, we employ the Arbitrary Lagrangian Eulerian (ALE) method\cite{donea2004arbitrary}. The velocity of the mesh, $\mathbf{v}_m$, at any point in the computational domain is obtained by solving a Laplace equation $\nabla^2 \mathbf{v}_m = \mathbf{0}$, with Dirichlet boundary conditions $\mathbf{v}_m = \mathbf{v} \cdot \mathbf{n}\mathbf{n} $ on the bubble boundary, $\Gamma_1$, and $\mathbf{v}_m = d z_\text{cm}/dt \, \, \hat{\mathbf{z}}$ on $\Gamma_2$. The boundary condition on $\Gamma_2$ moves the entire computational boundary upwards whenever the bubble center of mass is displaced. This condition minimizes mesh deformation and removes the need to remesh.
We use an implicit time-integration method with an adaptive second-order implicit Euler method. The resulting nonlinear system of equations is solved using the Newton method with a tolerance set to $10^{-8}$ and the direct solver PARDISO.

\begin{figure}
\includegraphics[width=0.48\textwidth]{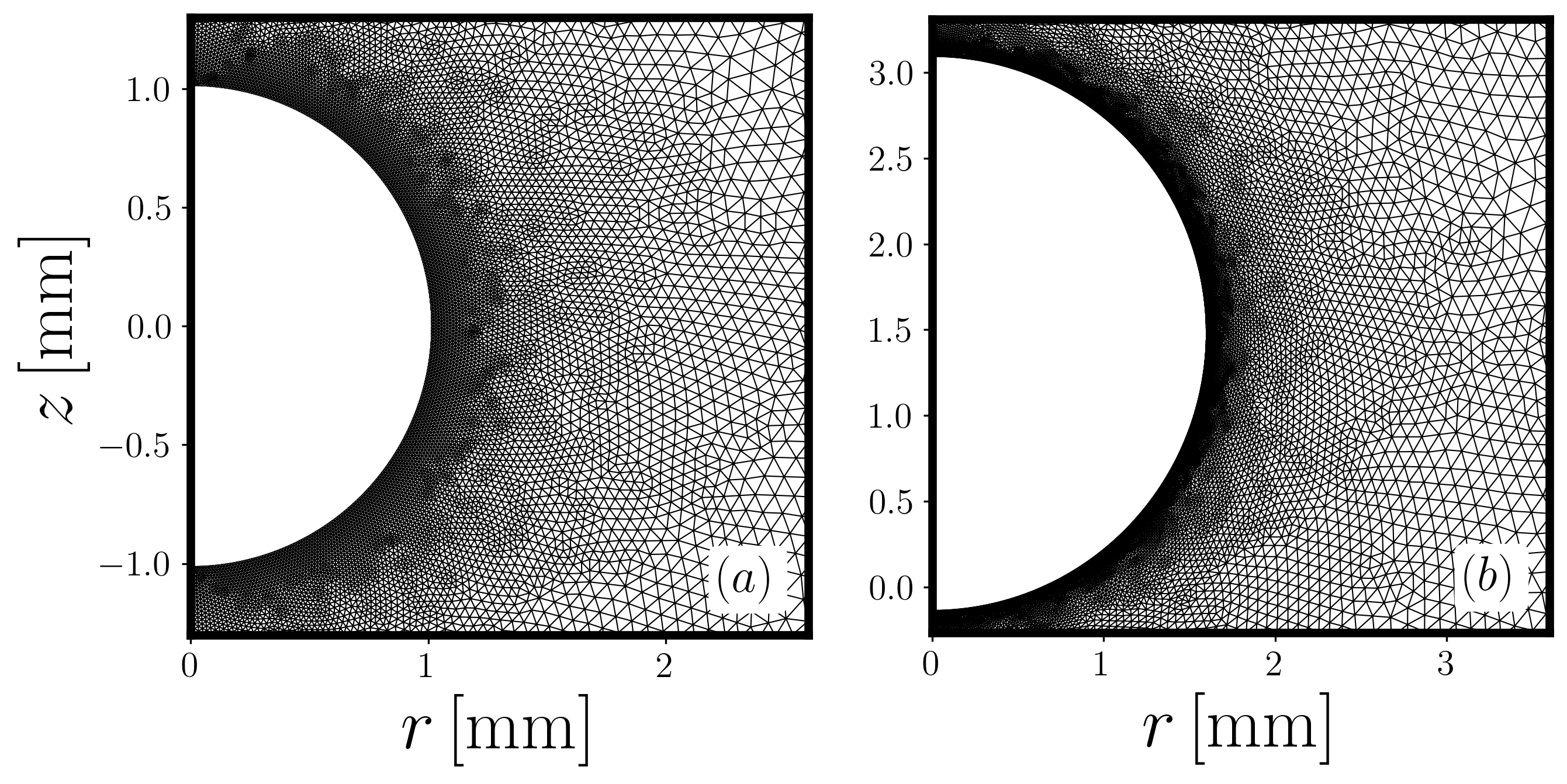}
\caption{\label{fig:Mesh_t0_tM} Zoom view of the mesh used for simulations. (a) Undeformed the mesh for a bubble of $R=1\rm{mm}$ at time $t=0$. (b) Deformed mesh for the maximum deformation. The plotted case is for a $k=0.75$ and for $f = 10 \ \text{Hz}$}
\end{figure}

\subsection{Characteristic dimensionless numbers}
To give a perspective of the physical regimes studied in this work, it is useful to compute the characteristic dimensionless numbers involved in this problem. 
The non-dimensional numbers associated with the purely buoyant rise of a bubble is given by the Bond number, defined as:
\begin{equation}
    Bo = \frac{\rho g R^2}{\sigma} \, .
\end{equation}
Using the parameters shown in Table~I, and the range of bubble radii $R \in [0.5, 2]\,\mathrm{mm}$, the Bond number spans the range $Bo \in [0.0324, 0.5190]$. As a consequence, the bubble equilibrium shape is  very close to a sphere. This agrees with the experimental observations~\cite{clift1978bubbles,iwata2008pressure}. 

The next question is how important are inertial effects in the problem under study. By considering the characteristic rising velocity of the bubble in the absence of pressure oscillations, it is possible to write:
\begin{equation}
    Re = \frac{\rho^2 g R_0^3}{\eta_0^2} \, .
\end{equation}
By plugging the values considered in this work, we obtain values of the Reynolds number typically around $Re \approx 10^{-7}-10^{-5}$, indicating that the rising regime in the absence of pressure oscillations is dominated by viscous forces. 

In the estimate of the Reynolds number we used the zero-shear viscosity. In fact, even in the absence of pressure oscillations, the shear-rate due to the natural bubble rising could be sufficient to trigger shear-thinning effects and significantly reduce the surrounding viscosity. This is a good approximation since the Carreau number $Cu$, given by 
\begin{equation}
    Cu = \frac{\rho g R \lambda}{\eta_0} \, ,
\end{equation}
is of order one. This implies that the characteristic rising velocity predicted by the Stokes law is a reasonable approximation but the viscosity could be somewhat smaller than the value of $\eta_0$ used in the estimate of the Reyndols number. Nevertheless, even a viscosity ten times smaller than $\eta_0$ would yield a Reynolds number smaller than $10^{-3}$, implying that inertial effects are negligible during the natural rising of the bubble. 

To verify this point, we performed numerical simulations of bubble rising in the absence of a pressure driving by considering a no-slip boundary condition on $\Gamma_2$. We computed the steady-state natural rising speed of the bubbles, $v_0$, which are reported in Table~II. In the case of $R=1\,\mathrm{mm}$, the rising velocities of the bubbles, $v_0$, are close to those predicted by the Stokes law for a bubble, $v_{\text{Stokes}} = \rho g R^2/ 3 \eta_0$~\cite{clift1978bubbles}. As the bubble becomes bigger, the natural rising velocity deviates by a factor of up 2.5 in the case of the largest bubble radius $R=2\,\mathrm{mm}$ due to increasingly large shear-thinning of the fluid surrounding the bubble. The values of the natural rising velocity of the bubbles, $v_0$, reported in Table~II will be used as a comparison for the rising speed of bubbles driven by pressure oscillations. 

\begin{table}[ht]
\caption{Rising velocities for three different bubbles in the absence of pressure oscillations.}
\label{tab:rise_velocities}
\centering
\begin{tabular}{ccc}
\hline
$R$ & $v_0$ [mm/s] & $v_{\text{Stokes}}$ [mm/s] \\
\hline
$0.5\,\mathrm{mm}$ & 0.0104 & 0.0091 \\
$1\,\mathrm{mm}$   & 0.0534 & 0.0363 \\
$2\,\mathrm{mm}$   & 0.4135 & 0.1453 \\
\hline
\end{tabular}
\end{table}

To verify if the natural rising could lead to deformations of the bubble due to viscous stresses, we compute the Capillary number $Ca$, given by
\begin{equation}
Ca = \frac{\rho g R^2}{\sigma} \, ,
\end{equation}
which turns out to be the same as the Bond number and, thus, smaller than one. 

In summary, in the absence of pressure oscillations, we expect the bubbles considered in this work to rise at a speed close to that predicted by the Stokes law and to remain close to its initial spherical shape. However, when the pressure oscillations are applied, the bubble experiences fast volumetric oscillations which considerably change the viscosity and its rising speed. Since we consider pressure oscillating at frequencies in the range of 1 to 300~Hz, the Carreau number based on the frequency $Cu_\omega$, given by 
\begin{equation}
    Cu_\omega = 2 \pi f \lambda \, , 
\end{equation}
lies between 300 and $10^5$. This implies that the volumetric oscillations of the bubble will lead to significant changes to the local viscosity even if their amplitude is small. In these conditions, we expect the rising velocity of the bubble to be highly unsteady, with an average value that differs significantly from that predicted by the Stokes law. Since an \textit{a priori} estimate of the average rising velocity is not possible, the importance of inertial and unsteady effects cannot be ruled out. For this reason, we will present the results using dimensional variables and we will discuss the importance of inertial and viscous forces once the kinematics of the bubble is computed.

\section{Validation of the code}
\begin{figure}[ht]
\centering
\includegraphics[width=\linewidth]{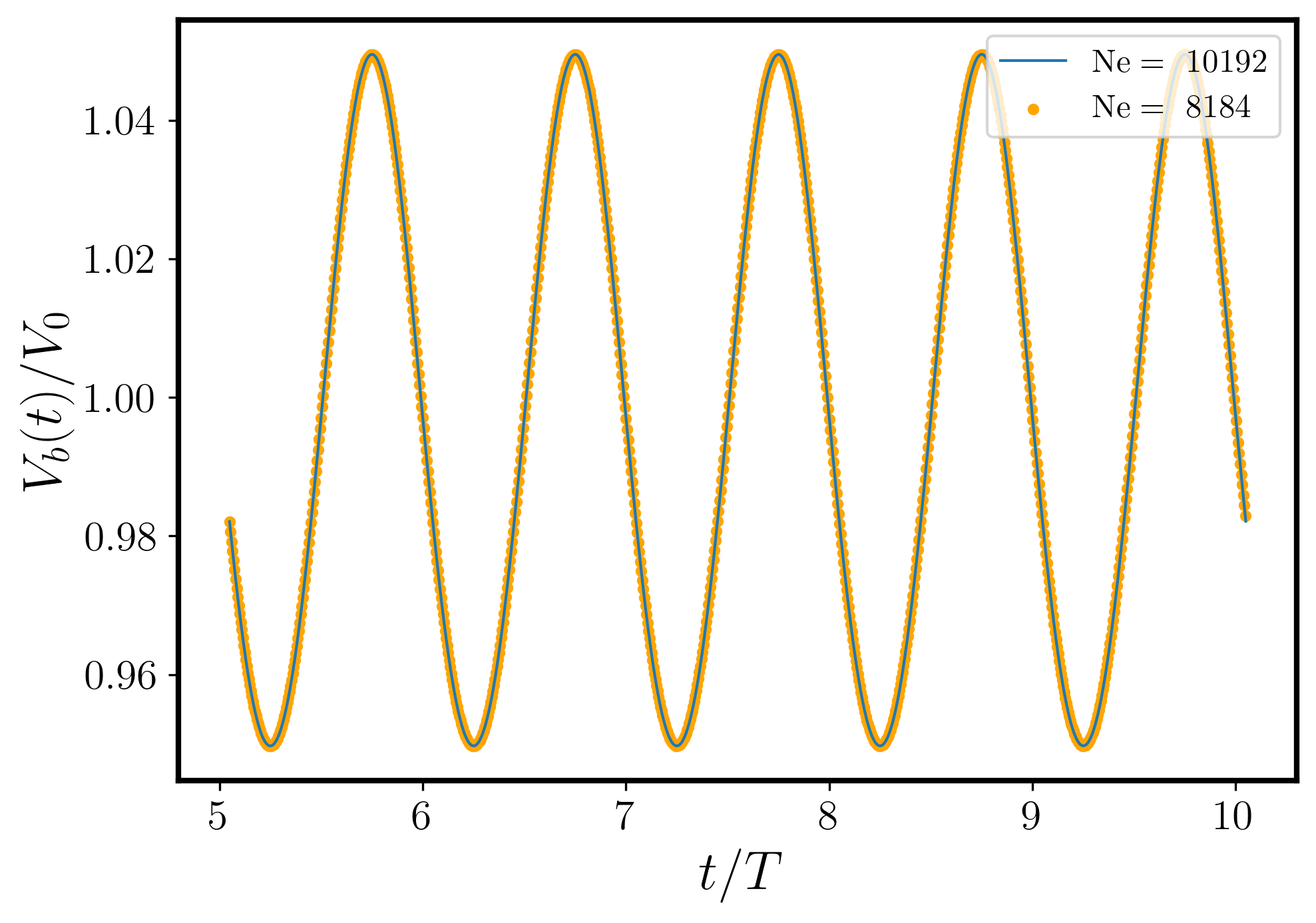}
\caption{Mesh convergence study for two different mesh resolutions. We consider the bubble volume normalized with respect its initial value as a mesh convergence metric in function of the time normalized with the period $T = 1 \ s$. Increasing the number of elements, $N_e$, by 2000 results in nearly identical curves, indicating that the results are independent of the mesh.} 
\label{fig:mesh_refinement}
\end{figure}

To validate the code and ensure the accuracy of the numerical results, we performed several mesh convergence tests. These tests confirmed that the selected mesh resolution shown in Figure \ref{fig:Mesh_t0_tM} is sufficient to capture the dynamics of the problem without introducing significant numerical errors.
Figure~\ref{fig:mesh_refinement} illustrates a representative mesh convergence study for a specific test case, characterized by the parameters $(k, f, R) = (0.05,\ 1~\text{Hz},\ 0.5~\text{mm})$. In this case, the number of elements in the computational domain, $N_e$, were increased from 8184 to 10192 by mostly refining the region close to the bubble surface where we expect the largest gradients. The plot shows the normalized bubble volume as a function of time, using the initial volume as a reference $V_B(t)/V_0$.
The results indicate that the solutions obtained with different mesh resolutions collapse onto each other, demonstrating that the results are independent of the mesh. Based on these findings, the mesh configuration used in Figure~\ref{fig:Mesh_t0_tM} was adopted for all subsequent simulations.

\section{Results}
\begin{figure}[ht]
\includegraphics[width=1.0\linewidth]{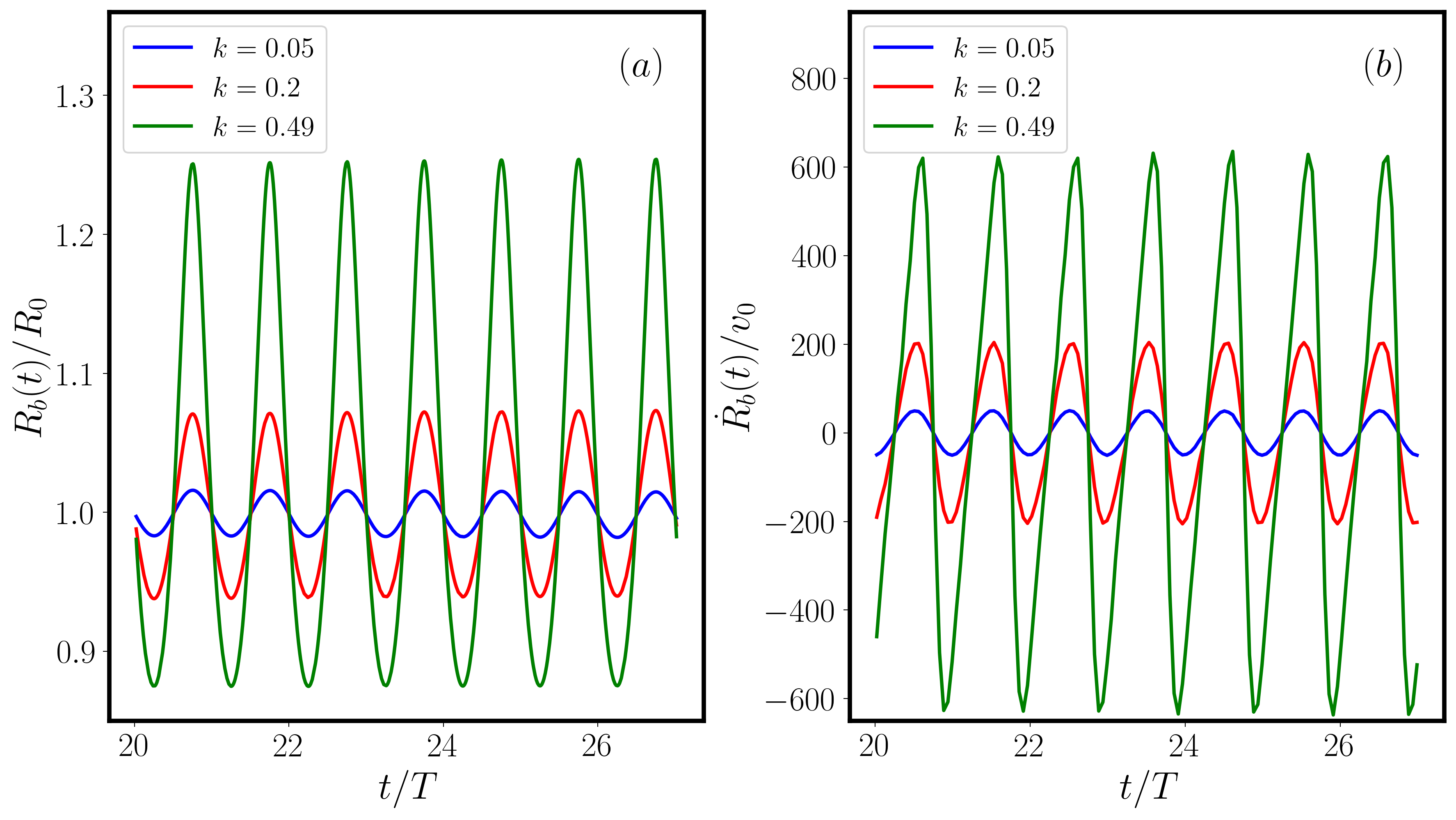}
\caption{Temporal evolution of the bubble radius and radial velocity for $f = 10~\text{Hz}$ and $R = 0.5~\text{mm}$. In panel (a), the radius of the bubble is plotted for different driving amplitudes. In panel (b), the radial velocity of the bubble, normalized by the bubble natural rising velocity, is shown. The time is normalized by the period of the external pressure driving $T=1/f$.}
\label{fig:bubble_radius}
\end{figure}

We begin our numerical investigation by studying bubbles of $R= 0.5 \rm{mm}$.
We apply a sinusoidal oscillating external pressure, as described in Equation \eqref{press_driv} with different amplitudes and frequencies. As a result of the external pressure driving, the bubble radius changes in time. In Figure~\ref{fig:bubble_radius}, we report the temporal evolution of the effective bubble radius \( R(t) \) for $f=10\rm{Hz}$ and the radial velocity at the bubble surface.
Since the bubble shape remains very close to that of a sphere, in both plots of Figure~\ref{fig:bubble_radius}, the radius is evaluated by taking the maximum value of the \( r \)-coordinate on the bubble boundary denoted as \( \Gamma_2 \), which corresponds to the horizontal radius of the bubble. The bubble velocity is taken as the time derivative of the radius. The results shown in panel (a) illustrate that the radial excursion of the bubble follows the frequency of the driving pressure and it increases with its amplitude. For all the cases considered, the radial excursion stays below 25\% of the initial bubble radius. For $k=0.75$, an asymmetry between the compressive and expansion of the bubble is apparent, with larger excursion during the expansion of the bubble compared to the compression phase. This is one of the signatures of the nonlinear behavior of the bubble dynamics (Figure~\ref{fig:bubble_radius}. In panel (b) of  Figure~\ref{fig:bubble_radius}, the radial velocity is shown normalized by the natural rising velocity $v_0$ (See Table II). This plot illustrates that the fluid velocity due to the radial excursion of the bubble is much larger than that of a bubble rising in the absence of the external pressure driving. 

\begin{figure*}[ht]\includegraphics[width=\textwidth]{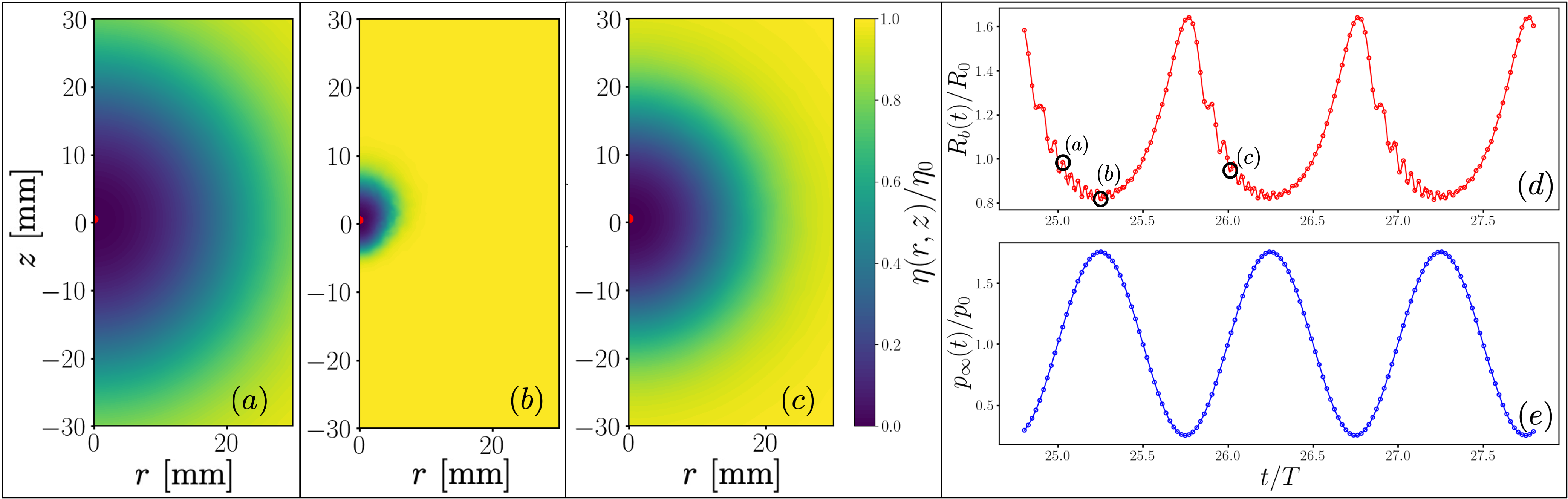}
\caption{Panels (a-c), snapshots of the normalized viscosity field with respect to the zero-shear viscosity. The bubble is shown in red, for comparison. Panel (d), radius of the bubble as a function of time, inside the plot are marked the radii values for each panel (a-c). Panel (e) external driving pressure as a function of time. The parameters used in these figures are $(f, k, R) = (300~\text{Hz}, 0.75, 0.5~\text{mm})$. }
\label{fig:viscosity_field}
\end{figure*}

In this case, it is apparent that the characteristic velocity and rate of strain around the bubble are dominated by the radial motion of the bubble surface in response to the external pressure driving. As a consequence, we expect the viscosity of the liquid to be drastically reduced compared to the case of a naturally rising bubble. 
To demonstrate the interplay of the radial oscillations and shear thinning, in Figure~\ref{fig:viscosity_field}(a-c), we show three snapshots of the viscosity field for a bubble of $R=0.5 \rm{mm}$ driven at $k=0.75$ and $f=300 \rm{Hz}$. The simulations reveal that the fast radial motion of the bubble introduces a very large region where the viscosity is greatly reduced compared to the zero-shear viscosity, $\eta_0$. 
These snapshots correspond to the points marked in Figure~\ref{fig:viscosity_field}(d-e). It is apparent that when the bubble is close to its equilibrium radius, its radial velocity is maximum, and the region where the viscosity is reduced is the largest. Conversely, when the bubble is close to the maximum or the minimum radii, then its radial velocity becomes close to zero, and the region where the viscosity thins becomes smaller.

\begin{figure*}[ht]
\includegraphics[width=\textwidth]{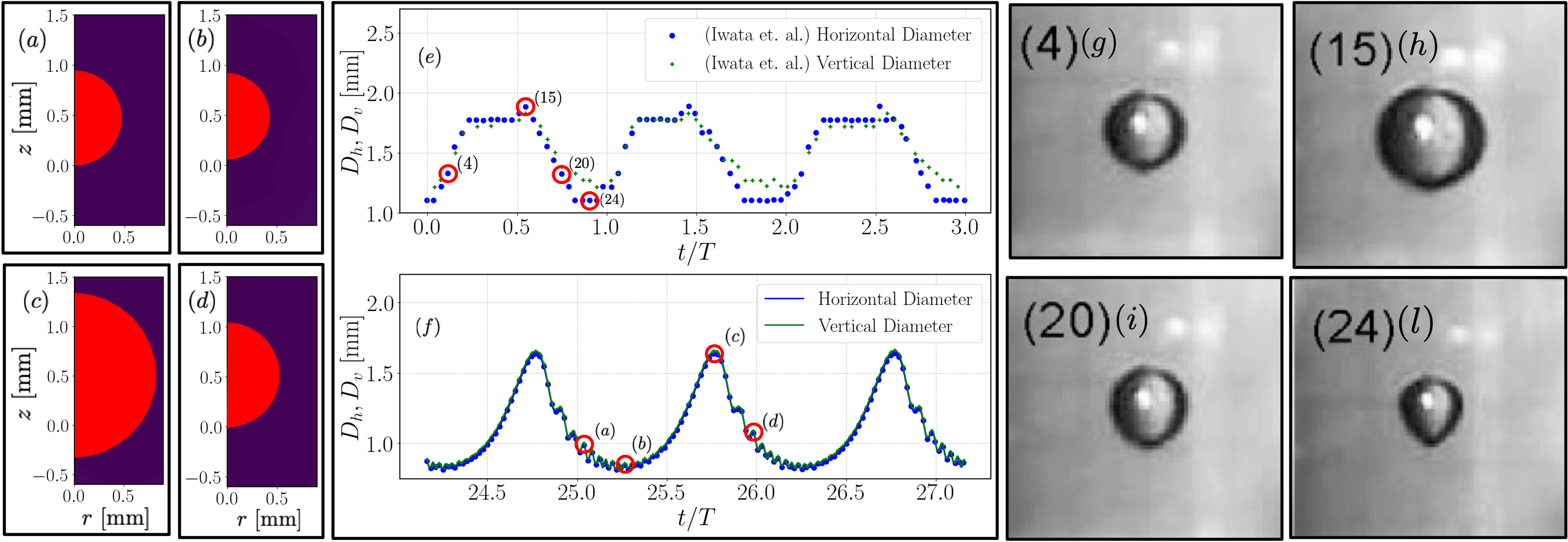}
\caption{Comparison between our numerical results and the experiments of Iwata \textit{et al}.\cite{iwata2008pressure} Panel (a) shows the horizontal and vertical measurements in the experiments. Panel (b) shows the 
same quantities obtained in the numerical simulations. Panels (e-h) show a sequence of bubble snapshots at different time instants for the case $(f, k, R_0) = (300~\text{Hz}, 0.75, 0.5~\text{mm})$. The right panels (g-l) show reference snapshots taken directly from Iwata \textit{et al.} \cite{iwata2008pressure}, with time indices corresponding to specific points in the graph (Reproduced with permission from J. non-Newton. Fluid Mech. 151, (1-3) (2008) Copyright Elsevier 2008).}
\label{fig:snapshots}
\end{figure*}

It is useful to compare our numerical results to those of Iwata \textit{et al}\cite{iwata2008pressure} at $f=300 \rm{Hz}$. In Figure~\ref{fig:snapshots}(e-f), we report the evolution of the bubble vertical and horizontal diameter measured by Iwata \textit{et al}. and in our simulations. In our simulations, the vertical diameter is computed as the difference between the largest and smallest z-coordinate of the bubble surface, and the horizontal diameter as twice the largest r-coordinate of the bubble surface (see Figure~\ref{fig:snapshots}(a-d)). We considered a 1mm diameter bubble, which is slightly smaller than that used in the experiments. Since in the experiments they do not control the pressure amplitude, we had to guess the value of $k$ that adjusts to the experiments. By using a value of $k=0.75$, we find that the excursions of the bubble vertical and horizontal diameters in the experiments and in the simulations are similar. In our simulations, the bubble remains almost spherical throughout the compression-expansion cycle. This is shown by the overlapping green and blue curves in Figure~\ref{fig:snapshots}(f) and it is also apparent in the snapshots Figure~\ref{fig:snapshots}(a-d). During most of the compression and expansion of the bubble, our numerical results are in qualitative agreement with the experiments, Figure \ref{fig:snapshots}(e-f)). However, close to the minimum radius of the bubble Figure \ref{fig:snapshots}(e)), the bubble develops a weak cusp at the rear, which is the signature of viscoelastic effects that are ignored in our model. The onset of the cusp is also apparent in Figure~\ref{fig:snapshots}(i), whereby the vertical and horizontal diameters in the experiments do not overlap anymore.

\begin{figure*}[ht]
\includegraphics[width=0.9\textwidth]{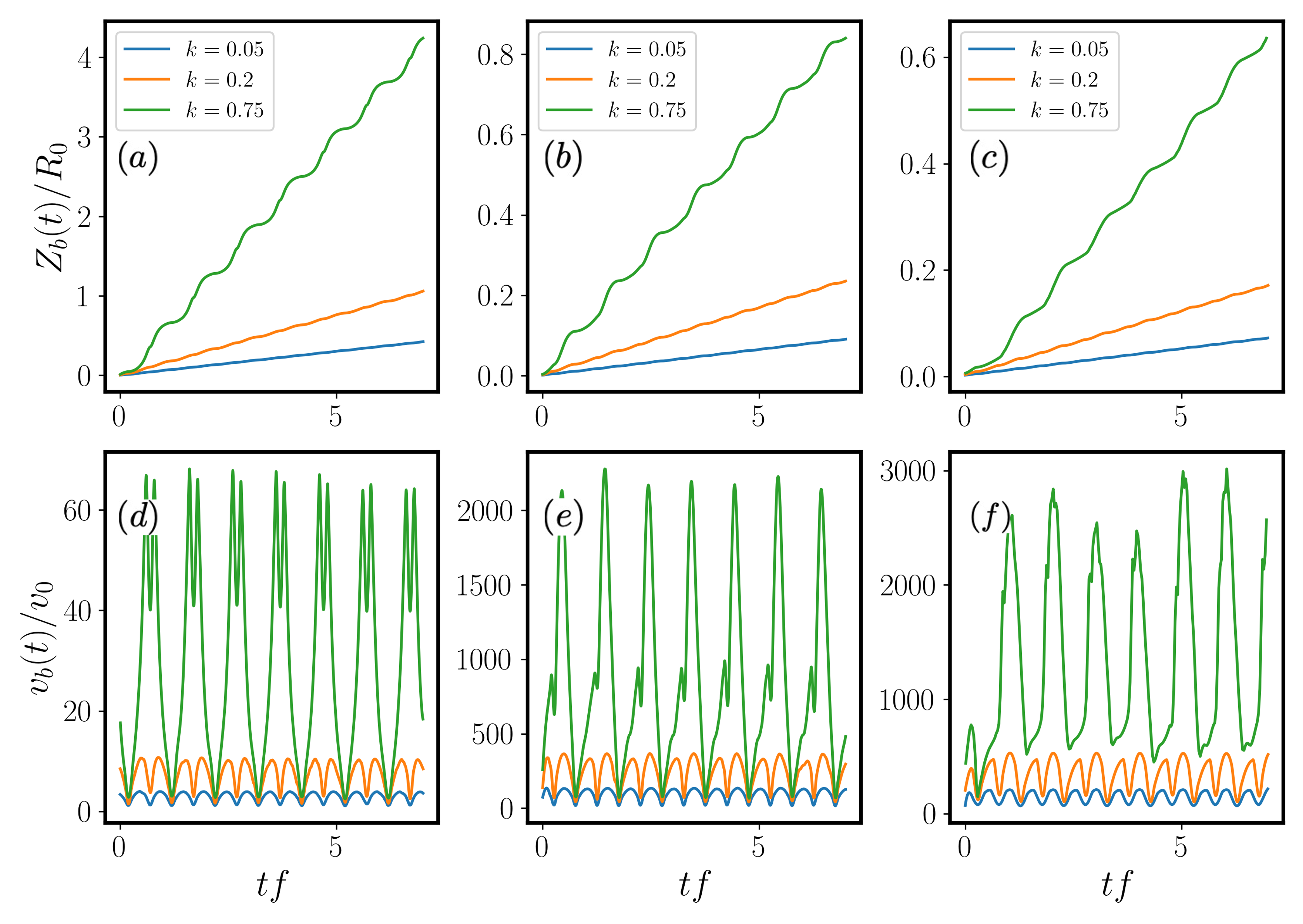}
\caption{
Kinematics of a bubble with initial radius $R_0 = 0.5~\text{mm}$ under different forcing amplitudes $k = 0.05, 0.2, 0.75$. Dimensionless time $t f$ is used, scaled with forcing frequencies $(f_1,f_4,f_5) = (1,150,300)~\text{Hz}$. 
Panels (a-c) show the dimensionless vertical position of the bubble $Z_b(t)/R_0$ for increasing forcing frequencies: (a) $f_1 = 1~\text{Hz}$, (b) $f_4 = 150~\text{Hz}$, and (c) $f_5 = 300~\text{Hz}$. 
Panels (d-f) report the corresponding dimensionless bubble velocity $v_b(t)/v_0$ for the same frequencies. 
Note that the vertical axes have different ranges in the three panels.
}

\label{fig:bubble_kinematics}
\end{figure*}

The time-dependent decrease in the viscosity surrounding the bubble shown in Figure \ref{fig:viscosity_field} results in a complex rising dynamics of the bubble. This is summarized in Figure~\ref{fig:bubble_kinematics} where we plot the dynamics of the bubble center of mass, $z_\text{cm}(t)$, and its rising velocity $v_b$ for a bubble of $R=0.5 \rm{mm}$. 
The plots in the first row depict the dynamics of the z-coordinate of the bubble’s center of mass in relation to time, normalized with the corresponding period. The position is determined using Equation \eqref{eq:pos} and is subsequently non-dimensionalized with the initial radius. At time $t=0$, the initial bubble radius is 0.5 mm. To provide a comprehensive overview while minimizing the number of plots, only the most representative frequencies and $k$-values were selected. Nonetheless, these plots capture the essential characteristics of the bubble’s behavior. The frequencies considered are $f=$1 Hz, 150 Hz, and 300 Hz.
For small-amplitude radial oscillations, the $z$-coordinate of the bubble center of mass, $Z_b(t)$, grows nearly linearly in time. By increasing the amplitude of the pressure driving, $Z_b(t)$ exhibits increased oscillations and also a larger mean velocity. This complex dynamics is further shown in the three plots in the bottom panels of Figure~\ref{fig:bubble_kinematics}. Here, the velocity—normalized with respect to the velocity of the bubble without pressure oscillations (see Tab.~\ref{tab:rise_velocities})—is obtained by applying a centered-difference numerical derivative of $Z_b(t)$. The plots reveal that the bubble rising velocity can transiently achieve values that are larger than its natural rising speed by up to three orders of magnitude. The transient rising velocity deviates significantly from a harmonic function, even at the smallest $k$ and $f$ studied here. It is apparent that, at a fixed frequency, increasing the pressure driving amplitude $k$ increases both the mean rising velocity and the oscillations around the mean. By comparing the results shown in panels (d-f) of Figure~\ref{fig:bubble_kinematics}, we observe that increasing the frequency of the pressure oscillations leads to an increase in the bubble rising speed. Note that the limits of the vertical axes are different in the panels (d-f).  
These simulations suggest that the fast radial motion, introduced by the volumetric oscillations of the bubble, and the ensuing shear thinning of the liquid have a strong impact on the rising dynamics of the bubble. As the driving pressure or its frequency increases, the expansion and contraction motions of the bubble become faster, resulting in a progressively more important shear thinning. This, in turn, results in a smaller frictional resistance and thus in a larger bubble rising speed.

\begin{figure}
\includegraphics[width=0.48\textwidth]{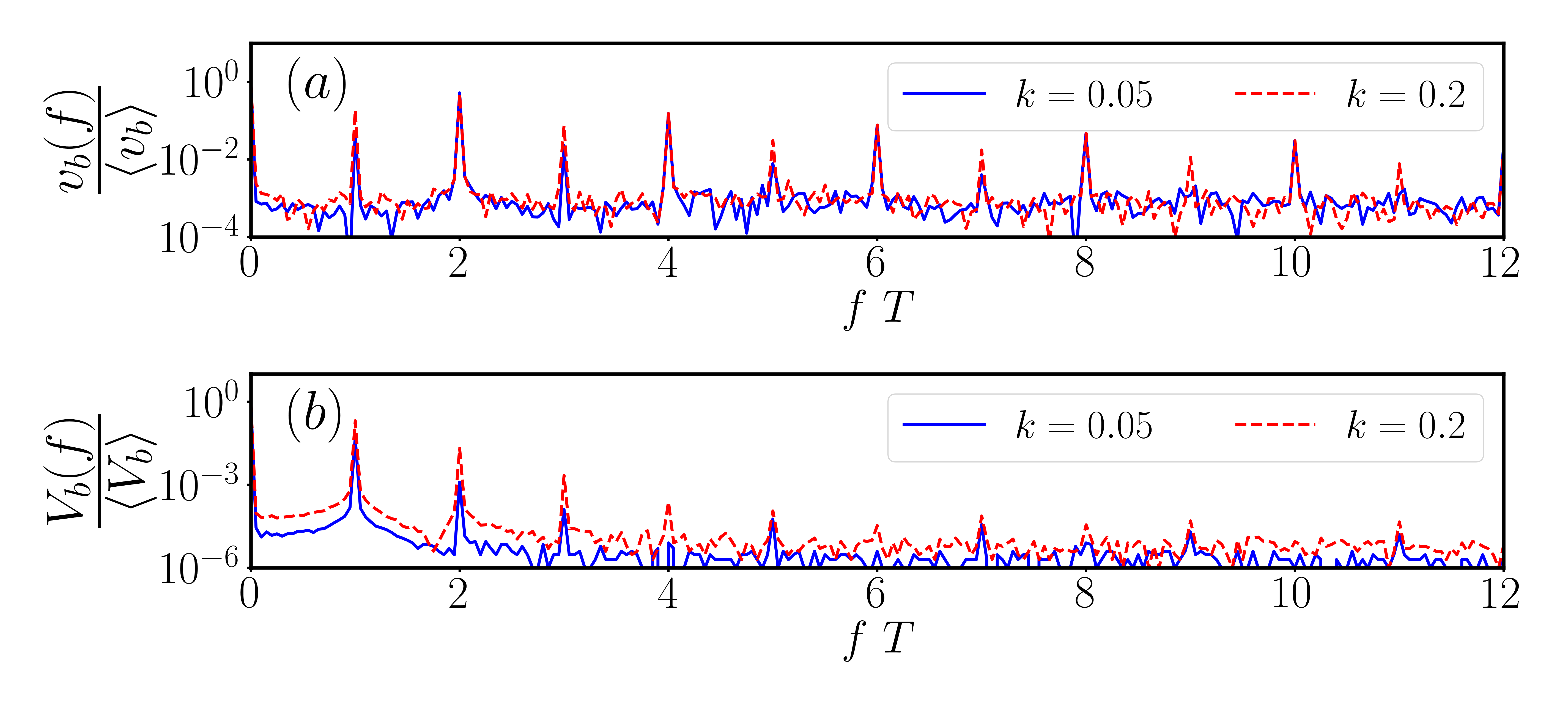}
\caption{Fourier analysis of (a) the velocity and (b) the volume of the bubble normalized by their mean value. We consider the cases of two amplitudes $k=0.05$ and $k=0.2$, with $R_0 = 0.5 \text{mm}$ and $f = 75 \text{Hz}$.}
\label{fig:FFT}
\end{figure}

In Figure~\ref{fig:bubble_kinematics}, we observe that the position and velocity plots exhibit highly non-linear dynamics with complex harmonics. Indeed, the velocity response does not follow a single harmonic, unlike the external pressure input. To fully appreciate the complexity of the rising speed dynamics,  we performed a Fourier analysis of the bubble dynamics using the MATLAB FFT algorithm. Specifically, in Figure~\ref{fig:FFT}, we display the frequency spectra of the volume and the velocity of the bubble. The spectra are plotted as a function of the frequency, non-dimensionalized with the period of the driving. The velocity and volume signals were normalized with respect to the mean value of the simulation.
In the upper plot of Figure~\ref{fig:FFT}, a strongly non-linear behavior of the rising velocity is observed, characterized by several peaks at multiples of the driving frequency. The magnitude of the peaks decreases slowly with increasing frequency, which means that higher-order harmonics are important in the transient response of the velocity. The second harmonic is the most prominent, and even harmonics display a larger magnitude than odd harmonics. In contrast, the bubble volume signal, shown in the lower plot of Figure~\ref{fig:FFT}, displays harmonics whose amplitude decays much faster with increasing frequency than those of the velocity. Since the plot is on a log-linear scale, the decay appears to be exponential with the harmonic number. The appearance of higher-order harmonics in the volume signal is a signature of nonlinear bubble dynamics due to large-amplitude oscillations and the complex rheology of the liquid\cite{de2019oscillations}.
Considering the spectra at two different amplitudes, it can be observed that they overlap at the same peak frequencies.

\begin{figure*}[ht]
\includegraphics[width=0.9\textwidth]{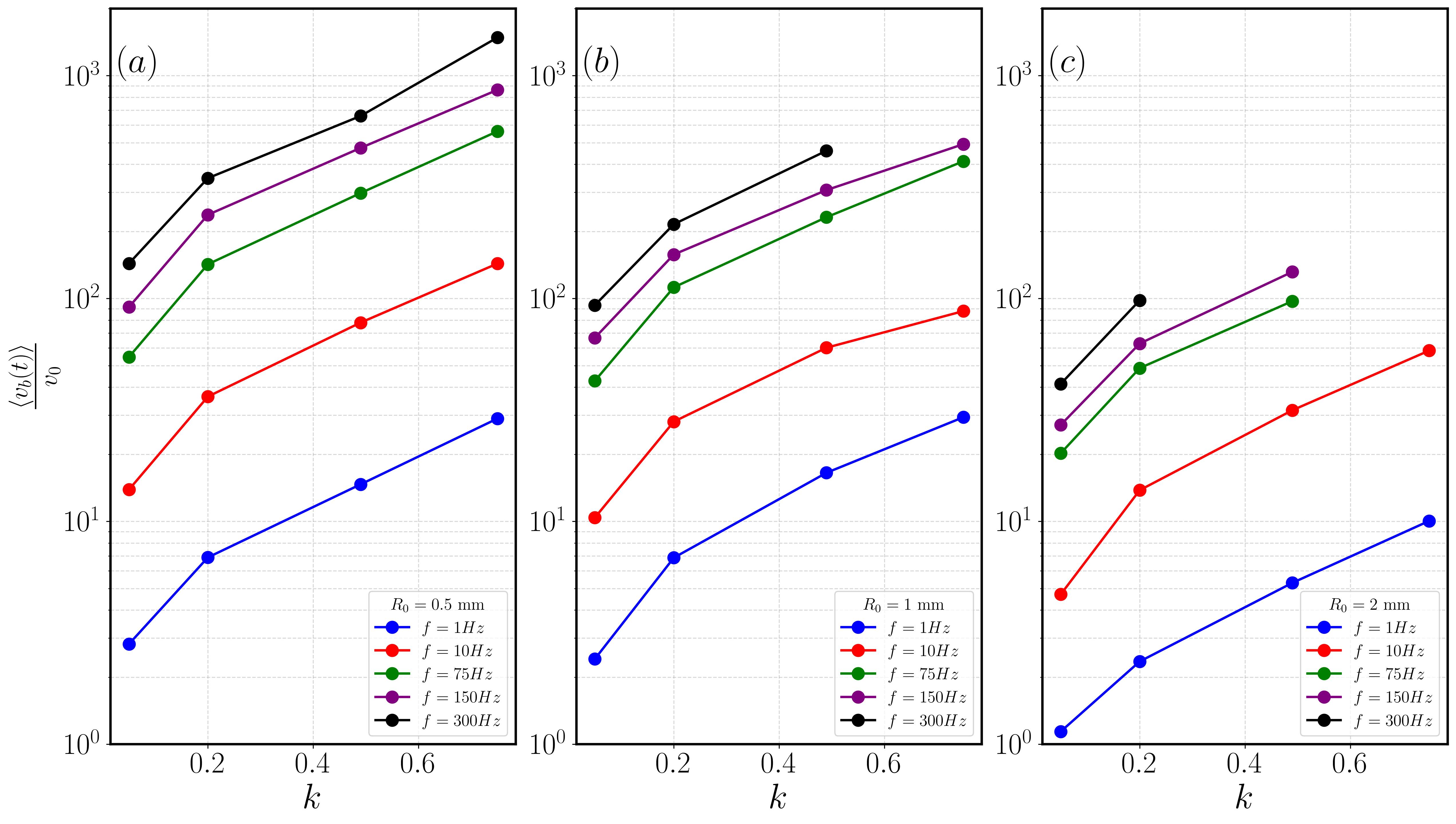}
\caption{Bubble rising speed averaged over one period for different initial radii $R_0=0.5$mm (a), $R_0=1$mm (b), and $R_0=2$ mm (c) plotted as a function of the pressure ampliture $k$ and for different frequencies $f$.}
\label{fig:summary_trend}
\end{figure*}

A comprehensive analysis of the simulations performed considered bubbles of three different bubble initial radii, $R_0=0.5, 1$, and $2$ mm, and pressure drivings at different $k$ and $f$ is shown in Figure~\ref{fig:summary_trend}. In this plot, the horizontal axis represents the amplitude of the pressure oscillation, $k$, and the vertical axis the mean rise velocity normalized with respect to their natural rising velocity, $k=0$. In Figure~\ref{fig:summary_trend}(b) and (c), some datapoints are missing. For these values of the parameters, the bubble develops a shape instability whereby the spherical shape is lost and the simulation eventually diverges. These shape instabilities have been observed experimentally at large driving pressures \cite{saint2020acoustic}. 
Across all of the radii investigated, increasing the amplitude and the frequency of the pressure driving enhances the mean rising velocity of the bubble. The effect, however, is strongly nonlinear: initial increases in amplitude or frequency lead to a marked increase in the mean velocity. Further increments result in diminishing returns, suggesting that additional energy input is not efficiently converted into a larger rising speed. This can be explained by the fact that the fluid viscosity can only thin up to a certain degree, and it approaches the solvent viscosity as the radial bubble oscillations become increasingly fast. 
The dependence of the normalized mean bubble rising velocity on the bubble radius is also interesting. Smaller bubbles exhibit the strongest relative gains compared to their natural rising velocity. This can be attributed to the fact that their rising speed in the absence of oscillations is too small to trigger any shear-thinning effect. Therefore, they experience the very large zero-shear viscosity during their natural rising (See Table II). Instead, larger bubbles of $R_0 =1 $ and $2$ mm, already experience some degree of shear-thinning during their natural rise (see Table II). Therefore, additional shear-thinning due to their radial oscillations is somewhat less effective in reducing the surrounding viscosity compared to the case of $R_0=0.5$mm.
Overall, the results indicate that the oscillatory pressure forcing is an effective mechanism to accelerate bubble transport in shear-thinning fluids, but its efficiency depends sensitively on both the forcing parameters and the bubble size, highlighting the inherently nonlinear nature of the process. 

\begin{figure*}
    \centering
    \includegraphics[width=1\linewidth]{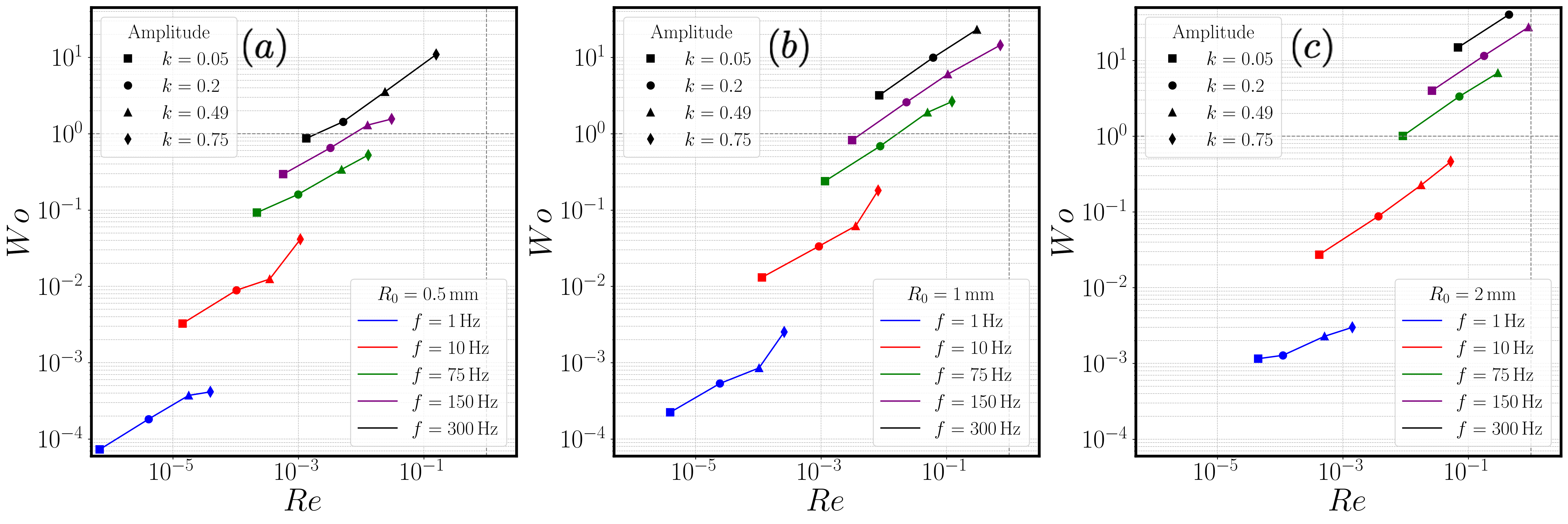}
    \caption{
Analysis of the dimensionless numbers: the Reynolds number $\mathrm{Re}$ (computed from \eqref{eq:av_Rey}) and the Womersley number $\mathrm{Wo}$ (computed from \eqref{eq:av_Wo}), shown as functions of forcing amplitude and frequency. Panels (a-c) correspond to fixed initial bubble radii $R_0 = 0.5\ \mathrm{mm}$ (a), $R_0 = 1\ \mathrm{mm}$ (b) and $R_0 = 2\ \mathrm{mm}$ (c), respectively. Each marker indicates a specific driving amplitude; different marker styles denote different driving frequencies (see legend).
}
    \label{fig:St_vs_Re}
\end{figure*}

We have seen that the bubble rising speed can increase by orders of magnitude compared to its natural rising velocity. At the same time, the viscosity of the fluid is also greatly reduced by the fast radial oscillations of the bubble. This suggests that inertial and unsteady effects in the Navier-Stokes equations, which are negligible for the natural rising case (see Section IIA), could become relevant when the oscillatory pressure driving is applied to the bubble. 
To investigate this point, we computed the mean Reynolds number and the mean Womersley number defined as
\begin{equation}
\langle Re \rangle = \frac{\rho \langle v_b \rangle R_0}{\langle \eta \rangle} \, \, ,
\label{eq:av_Rey}
\end{equation}
\begin{equation}
\langle Wo \rangle = \frac{ \, 2 \, \pi \, f \, \rho \,  R_0^2}{\langle \eta \rangle} \, \, ,
\label{eq:av_Wo}
\end{equation}
where $\langle v_b \rangle$ denotes the mean bubble rising speed and $\langle \eta \rangle$ denotes the mean viscosity at the surface of the bubble averaged over one period. The values of these two dimensionless numbers are shown in Figure~\ref{fig:St_vs_Re} for each of the simulations performed. It is apparent that unsteady effects, which were neglected in previous works \cite{de2019oscillations}, can become important. Even the smallest bubbles reach values of $\langle Wo \rangle$ of order one if driven at a sufficiently large frequency. Interestingly, these frequencies correspond to those used by Iwata \textit{et al}. \cite{iwata2008pressure} in their experiments. This finding suggests that the shear viscosity is sufficiently reduced due to shear thinning that the diffusion time of the vorticity around the bubble becomes comparable to the oscillation period. On the other hand, advective effects due to the rising velocity are typically negligible. The average Reynolds number only becomes of order one for bubbles of $R_0=1$mm and $R_0=2$mm at the largest $k$ and $f$. Overall, these findings suggest that inertial and unsteady effects might play a role in the experiments of Iwata \textit{et al}.\cite{iwata2008pressure}despite the very large zero shear viscosity displayed by their fluid.

\subsection{Comparison with Iwata Experiments}

\begin{figure}[ht]
\centering
\includegraphics[width=0.5\textwidth]{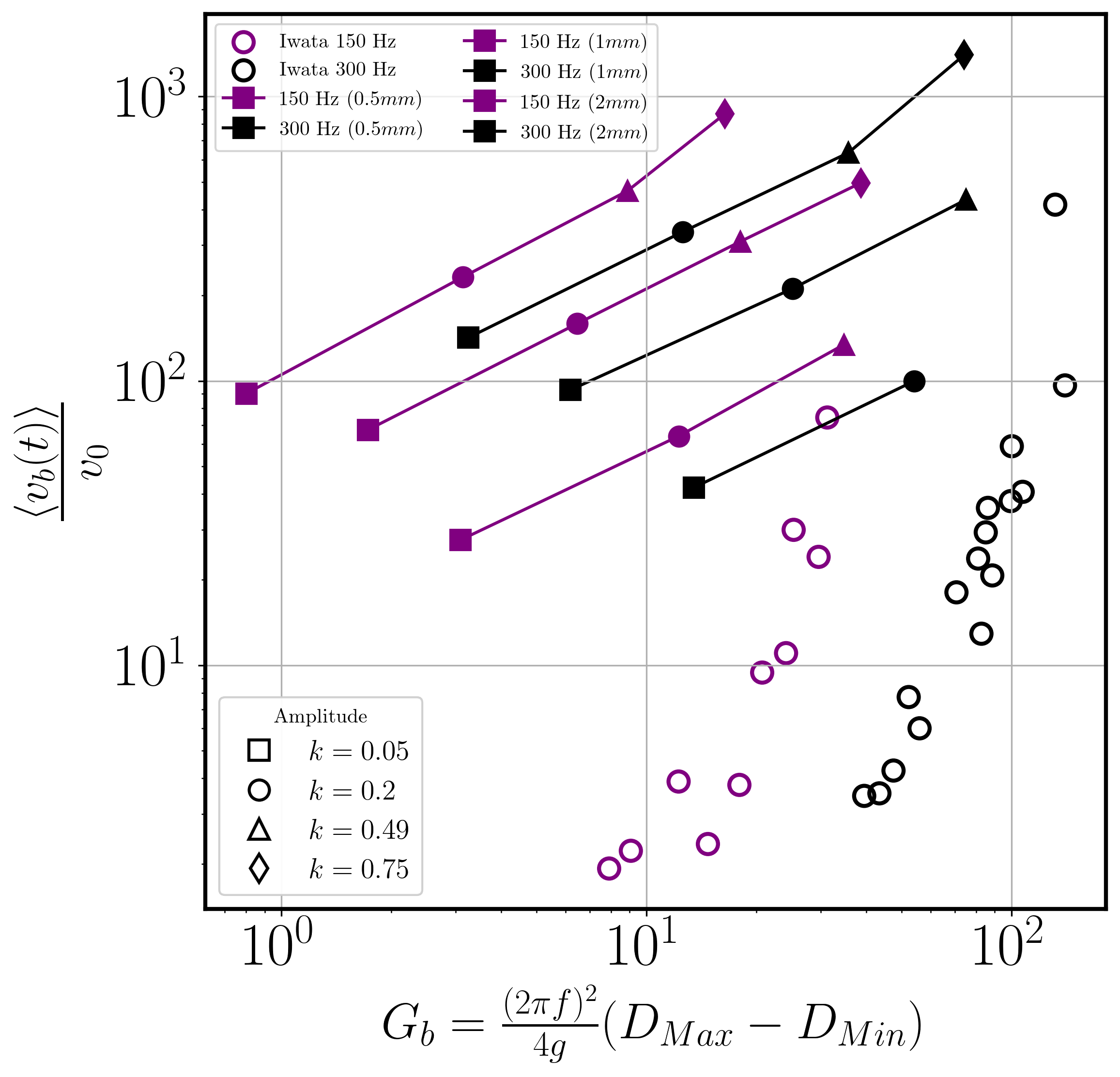}
\caption{Comparison of simulated and experimental results. The values in millimeters in the legend refer to the initial radius of the bubble in the simulations. The bottom axis was introduced by Iwata \textit{et al.}~\cite{iwata2008pressure} and represents a characteristic radial acceleration of the bubble surface. Here $D_{\text{max}}$ and $D_{\text{min}}$ are the maximum and minimum diameters that the bubble reaches during the rise. The experimental data points are extracted from Figure~8 of Iwata \textit{et al.}~\cite{iwata2008pressure}. }
\label{fig:iwata_comparison}
\end{figure}

Previous experimental studies \cite{iwata2008pressure} have focused on the rising of millimetric bubbles subjected to a periodically oscillating pressure. In this study, we compare our simulation results with these experimental findings, focusing on bubble shape and terminal rise velocity.  
The reference study conducted experiments within a system referred to as POD (Pressure Oscillation Defoaming), consisting of a quartz cell filled with the sample liquid and sealed with a rubber diaphragm cap. A vibrating piston, driven by a sinusoidal generator, induced periodic pressure oscillations inside the cell, causing the bubble to undergo repeated compression and expansion. These oscillations generated local radial flows, which, for shear-thinning fluids, reduced the apparent viscosity around the bubble and enhanced its rise velocity. The rising dynamics of the bubbles were analyzed as a function of a non-dimensional radial acceleration of their surface, $G_b$, defined as
\[
G_b = \frac{(2\pi f)^2 \bigl(D_{\text{max}} - D_{\text{min}}\bigr)}{4 g},
\]
where $D_{\text{max}}$ and $D_{\text{min}}$ are the maximum and minimum diameters attained by the bubble during one driving period, respectively. The dimensions of their bubbles are comparable to those considered in the present study. Figure~\ref{fig:iwata_comparison} shows their experimental data overlaid with the numerical results, where the x-axis represents the non-dimensional acceleration and the y-axis represents the velocity normalized by the no-oscillation case.  

The qualitative trends in the simulations align with those observed experimentally. However, significant quantitative discrepancies remain, suggesting that the current modeling approach may not fully capture the physical realities of the experiments. Nevertheless, the simulations effectively reproduce the general phenomena of enhanced bubble rising dynamics under oscillatory forcing.  

The study by Iwata \textit{et al.} \cite{iwata2008pressure} did not report the amplitude of the oscillations imposed on the system, and our investigation does not identify a value capable of fitting their data. In addition, we also observe discrepancies with the experiments in the bubble shape during a driving period and in the trend of the average rising velocity as a function of the acceleration. The primary reason may be that a shear-thinning model alone cannot capture the full rheology, since the fluid used in their experiments may exhibit elastic effects. Elastic stresses combined with shear thinning can explain the formation of the cusp observed in Figure~\ref{fig:snapshots} \cite{tsamopoulos2008steady,yuan2020dynamics}. 
Further simulations will be conducted considering constitutive models that account for fluid elasticity.

\section{Conclusions}

Using numerical simulations, we investigated the rising dynamics of millimeter-sized bubbles under external pressure oscillations within an incompressible, shear-thinning fluid. 
Motivated by previous experiments \cite{iwata2008pressure}, we employed the Carreau--Yasuda constitutive equation to model the shear-thinning properties of the fluid and we considered bubbles with an initial radius between $0.5\,\mathrm{mm}$ and $2\,\mathrm{mm}$.
We assumed an axisymmetric geometry and we neglected mass and heat transport between the gas and the fluid. We also neglected the viscosity of the gas, thus avoiding solving for the flow within the bubble. We employed an isothermal compression law to describe the gas pressure within the bubble.
 
Before studying the effect of pressure oscillations, we investigated the natural rising motion of the bubbles. We found that bubbles smaller than $1\,\mathrm{mm}$ closely follow Stokes law. Instead, bubbles approaching a $2\,\mathrm{mm}$ radius begin to significantly modify the local viscosity, leading to an increased rising velocity compared to that predicted by Stokes law.
We then applied an external pressure driving and studied the bubble rising and shape dynamics. To explore a wide range of cases comparable to those in the experiments of Iwata~et~al.~\cite{iwata2008pressure}, we changed the frequency and the amplitude of the periodic pressure driving.
The external pressure oscillations drive periodic bubble volume oscillations. This results in the appearance of large local deformation rates, which reduce the effective viscosity in the bubble’s surroundings. As a consequence, the bubble drag decreases and its rising velocity increases by several orders of magnitude compared to that in the absence of the external driving. 

The bubble remained almost spherical in all simulations, exhibiting pronounced nonlinear radial and rising dynamics. A Fourier analysis of the bubble kinematics confirms the presence of strong nonlinearities, which grow with increasing forcing frequency and amplitude. The radial dynamics is characterized mainly by the first multiples of the driving frequency. Instead, the rising velocity displays contributions from many higher-order harmonics, with larger contributions from even harmonics. Post-processing analysis of the dimensionless numbers confirms that the simulations were conducted in viscous regimes. However, the significant reduction of the shear viscosity can lead to effective Womersley and Reynolds numbers close to unity. This suggests that, even if inertial effects are absent during natural bubble rising, they can become relevant as external pressure oscillations are applied. 

Comparison with experimental data~\cite{iwata2008pressure} revealed a good qualitative agreement, particularly in terms of the observed enhancement in rising velocity. The results are not far from those reported by Iwata \textit{et al.}, but the trend as the oscillation amplitude is increased does not match that observed in the experiments. Notable discrepancies were also found in the bubble shape throughout the oscillation cycle. We speculate that these differences stem from the absence of an elastic component in the current model, which prevents capturing the full viscoelastic behavior of the experimental system. The mismatch in velocity is also attributed to the lack of elastic effects that can slow down the bubble and modify its rising path.

The main findings may be summarized by the following points:
\begin{itemize}
    \item Pressure oscillations induce strong volumetric bubble oscillations, generating large local strain rates in the surrounding fluid.
    \item The shear-thinning effect significantly reduces the effective viscosity near the bubble, leading to pronounced drag reduction.
    \item The bubble rising velocity increases by several orders of magnitude compared to natural rising, especially at higher forcing amplitudes and frequencies.
    \item Bubble dynamics are highly nonlinear, with the rising velocity exhibiting multiple harmonics and strong deviations from sinusoidal behavior.
    \item Smaller bubbles experience the largest relative enhancement in rising speed, as their natural motion does not trigger shear thinning.
    \item Numerical simulations qualitatively agree with experiments, but discrepancies in bubble shape and rise velocity suggest the need for viscoelastic constitutive models to fully capture the physical behavior of the reference work.
\end{itemize}

Future work should include fluid elasticity to bridge the gap with experiments and refine the quantitative predictions. At the same time, the present results provide a solid basis for both theoretical developments and potential applications aimed at improving bubble removal processes in industry without resorting to costly methods.

\section{Data Availability Statement}
The data that supports the findings of this study are available within the article.

\begin{acknowledgments}
This work was supported by the Ramon y Cajal fellowship RYC2021-030948-I and by the PID2022-139803NB-I00 research grant funded by the MICIU/AEI /10.13039/501100011033 and by the EU under the NextGenerationEU/PRTR program. Mario Riccio was supported by the FPI PhD fellowship PRE2023-UZ-10 funded by MICIU/AEI /10.13039/501100011033.
The authors thank Dr.~Paula Mart\'{\i}nez Lera for the help with the Fourier analysis. 
\end{acknowledgments}

$\,$

$\,$


\begin{thebibliography}{99}

\bibitem{iwata2008pressure}
Iwata, S., Yamada, Y., Takashima, T., \& Mori, H. (2008).
Pressure-oscillation defoaming for viscoelastic fluid.
\textit{J. Non-Newton. Fluid Mech.}, 151(1–3), 30–37. Elsevier.

\bibitem{eskin2017overview}
Eskin, D. G. (2017).
Overview of ultrasonic degassing development.
In \textit{Light Met. 2017} (pp. 1437–1443). Springer.

\bibitem{clift1978bubbles}
Clift, R., Grace, J. R., \& Weber, M. E. (1978).
\textit{Bubbles, Drops, and Particles}.
Academic Press, New York.

\bibitem{Pillapakkam2007}
Pillapakkam, S. B., Singh, P., Blackmore, D., \& Aubry, N. (2007).
Transient and steady state of a rising bubble in a viscoelastic fluid.
\textit{J. Fluid Mech.}, 589, 215–252.

\bibitem{PILZ2007124}
Pilz, C., \& Brenn, G. (2007).
On the critical bubble volume at the rise velocity jump discontinuity in viscoelastic liquids.
\textit{J. Non-Newton. Fluid Mech.}, 145(2), 124–138. Elsevier.

\bibitem{fakhari2023single}
Fakhari, A., \& Fernandes, C. (2023).
Single-bubble rising in shear-thinning and elastoviscoplastic fluids using a geometric volume of fluid algorithm.
\textit{Polymers}, 15(16), 3437. MDPI.

\bibitem{chen2022rise}
Chen, Q., Restagno, F., Langevin, D., \& Salonen, A. (2022).
The rise of bubbles in shear thinning viscoelastic fluids.
\textit{J. Colloid Interface Sci.}, 616, 360–368. Elsevier.

\bibitem{tsamopoulos2008steady}
Tsamopoulos, J., Dimakopoulos, Y., Chatzidai, N., Karapetsas, G., \& Pavlidis, M. (2008).
Steady bubble rise and deformation in Newtonian and viscoplastic fluids and conditions for bubble entrapment.
\textit{J. Fluid Mech.}, 601, 123–164. Cambridge University Press.

\bibitem{fraggedakis2016velocity}
Fraggedakis, D., Pavlidis, M., Dimakopoulos, Y., \& Tsamopoulos, J. (2016).
On the velocity discontinuity at a critical volume of a bubble rising in a viscoelastic fluid.
\textit{J. Fluid Mech.}, 789, 310–346. Cambridge University Press.

\bibitem{stein2000rise}
Stein, S., \& Buggisch, H. (2000).
Rise of pulsating bubbles in fluids with a yield stress.
\textit{Z. Angew. Math. Mech.}, 80(11–12), 827–834. Wiley.

\bibitem{de2019rising}
De Corato, M., Dimakopoulos, Y., \& Tsamopoulos, J. (2019).
The rising velocity of a slowly pulsating bubble in a shear-thinning fluid.
\textit{Phys. Fluids}, 31(8). AIP Publishing.

\bibitem{de2019oscillations}
De Corato, M., Saint-Michel, B., Makrigiorgos, G., Dimakopoulos, Y., Tsamopoulos, J., \& Garbin, V. (2019).
Oscillations of small bubbles and medium yielding in elastoviscoplastic fluids.
\textit{Phys. Rev. Fluids}, 4(7), 073301. APS.

\bibitem{zhang2023drag}
Zhang, X., Sugiyama, K., \& Watamura, T. (2023).
Drag force on an oscillatory spherical bubble in shear-thinning fluid.
\textit{J. Fluid Mech.}, 959. Cambridge University Press.

\bibitem{gardner2023bubble}
Gardner, J. E., Wadsworth, F. B., Carley, T. L., Llewellin, E. W., Kusumaatmaja, H., \& Sahagian, D. (2023).
Bubble formation in magma.
\textit{Annu. Rev. Earth Planet. Sci.}, 51, 131–154.

\bibitem{namiki2016sloshing}
Namiki, A., Rivalta, E., Woith, H., \& Walter, T. R. (2016).
Sloshing of a bubbly magma reservoir as a mechanism of triggered eruptions.
\textit{J. Volcanol. Geotherm. Res.}, 320, 156–171. Elsevier.

\bibitem{guo2022characteristic}
Guo, T., et al. (2022).
Characteristic analysis of air bubbles on the rheological properties of cement mortar.
\textit{Constr. Build. Mater.}, 316, 125812. Elsevier.

\bibitem{gallego2015ultrasonic}
Gallego-Juárez, J. A., et al. (2015).
Ultrasonic defoaming and debubbling in food processing and other applications.
In \textit{Power Ultrasonics} (pp. 793–814). Elsevier.

\bibitem{iwata2019local}
Iwata, S., et al. (2019).
Local flow around a tiny bubble under a pressure-oscillation field in a viscoelastic worm-like micellar solution.
\textit{J. Non-Newton. Fluid Mech.}, 263, 24–32. Elsevier.

\bibitem{donea2004arbitrary}
Donea, J., et al. (2004).
Arbitrary Lagrangian Eulerian methods.
\textit{Encycl. Comput. Mech.}. Wiley.

\bibitem{astarita1965motion}
Astarita, G., \& Apuzzo, G. (1965).
Motion of gas bubbles in non-Newtonian liquids.
\textit{AIChE J.}, 11(5), 815–820. Wiley.

\bibitem{karapetsas2019dynamics}
Karapetsas, G., et al. (2019).
Dynamics and motion of a gas bubble in a viscoplastic medium under acoustic excitation.
\textit{J. Fluid Mech.}, 865, 381–413. Cambridge University Press.

\bibitem{lin1970mechanisms}
Lin, T. J. (1970).
Mechanisms and control of gas bubble formation in cosmetics.
\textit{J. Soc. Cosmet. Chem.}, 22(6), 323–337.

\bibitem{Taki2006BubCoal}
Taki, K., et al. (2006).
Bubble coalescence in foaming process of polymers.
\textit{Polym. Eng. Sci.}, 46(5), 680–690.

\bibitem{garstecki2005formation}
Garstecki, P., et al. (2005).
Formation of bubbles and droplets in microfluidic systems.
\textit{Bull. Pol. Acad. Sci. Tech. Sci.}, 361–372.

\bibitem{sofjan2004effects}
Sofjan, R. P., \& Hartel, R. W. (2004).
Effects of overrun on structural and physical characteristics of ice cream.
\textit{Int. Dairy J.}, 14(3), 255–262. Elsevier.

\bibitem{luyten2004crispy}
Luyten, H., et al. (2004).
Crispy/crunchy crusts of cellular solid foods: a literature review with discussion.
\textit{J. Texture Stud.}, 35(5), 445–492. Wiley.

\bibitem{yuan2020dynamics}
Yuan, W., et al. (2020).
Dynamics and deformation of a three-dimensional bubble rising in viscoelastic fluids.
\textit{J. Non-Newton. Fluid Mech.}, 285, 104408. Elsevier.

\bibitem{seropian2021review}
Seropian, G., et al. (2021).
A review framework of how earthquakes trigger volcanic eruptions.
\textit{Nat. Commun.}, 12(1), 1004.

\bibitem{seropian2023effect}
Seropian, G., et al. (2023).
The effect of mechanical shaking on the rising velocity of bubbles in high-viscosity shear-thinning fluids.
\textit{J. Geophys. Res. Solid Earth}, 128(5), e2022JB025741. Wiley.

\bibitem{garbin2025bubbles}
Garbin, V., et al. (2025).
Bubbles and bubbly flows.
\textit{Int. J. Multiph. Flow}, 105240. Elsevier.

\bibitem{lohse2003bubble}
Lohse, D. (2003).
Bubble puzzles.
\textit{Phys. Today}, 56(2), 36–41. AIP Publishing.

\bibitem{lohse2018bubble}
Lohse, D. (2018).
Bubble puzzles: from fundamentals to applications.
\textit{Phys. Rev. Fluids}, 3(11), 110504. APS.

\bibitem{dollet2019bubble}
Dollet, B., Marmottant, P., \& Garbin, V. (2019).
Bubble dynamics in soft and biological matter.
\textit{Annu. Rev. Fluid Mech.}, 51, 331–355.

\bibitem{hassager1979negative}
Hassager, O. (1979).
Negative wake behind bubbles in non-Newtonian liquids.
\textit{Nature}, 279(5712), 402–403.

\bibitem{pourzahedi2022flow}
Pourzahedi, A., et al. (2022).
Flow onset for a single bubble in a yield-stress fluid.
\textit{J. Fluid Mech.}, 933, A21. Cambridge University Press.

\bibitem{moschopoulos2021concept}
Moschopoulos, P., et al. (2021).
The concept of elasto-visco-plasticity and its application to a bubble rising in yield stress fluids.
\textit{J. Non-Newton. Fluid Mech.}, 297, 104670. Elsevier.

\bibitem{saint2020acoustic}
Saint-Michel, B., \& Garbin, V. (2020).
Acoustic bubble dynamics in a yield-stress fluid.
\textit{Soft Matter}, 16(46), 10405–10418. RSC.

\bibitem{pelekasis2004secondary}
Pelekasis, N. A., et al. (2004).
Secondary Bjerknes forces between two bubbles and the phenomenon of acoustic streamers.
\textit{J. Fluid Mech.}, 500, 313–347. Cambridge University Press.

\bibitem{rodriguez2015generation}
Rodríguez-Rodríguez, J., et al. (2015).
Generation of microbubbles with applications to industry and medicine.
\textit{Annu. Rev. Fluid Mech.}, 47, 405–429.

\bibitem{esposito2024buoyancy}
Esposito, G., Dimakopoulos, Y., \& Tsamopoulos, J. (2024).
Buoyancy induced motion of a Newtonian drop in elastoviscoplastic materials.
\textit{J. Rheol.}, 68(5), 815–835. AIP Publishing.


\end{thebibliography}
\end{document}